\documentclass[10pt,twocolumn,twoside]{IEEEtran}
\usepackage{amsmath,graphicx}
\usepackage[latin1]{inputenc}
\usepackage{psfrag}
\usepackage{color}
\usepackage{amssymb}
\usepackage{amsthm}
\usepackage{subfigure}
\usepackage{cite}

\newtheorem{theorem}{Theorem}
\newtheorem{corollary}{Corollary}

\title{Secrecy Outage Probability of Network-Coded Cooperative Communication}
\author{João~Luiz~Rebelatto,~\IEEEmembership{Member,~IEEE,} Richard~Demo~Souza,~\IEEEmembership{Senior~Member,~IEEE,} Rodrigo~Tsuneyoshi~Kaido,~\IEEEmembership{} Ohara~Kerusauskas~Rayel,~\IEEEmembership{Student~Member,~IEEE} and
Bartolomeu~F.~Uchôa-Filho,~\IEEEmembership{Senior~Member,~IEEE}%
\thanks{This work has been supported in part by CNPq and CAPES, (Brazil).}
\thanks{João Luiz Rebelatto, Richard Demo Souza, Rodrigo Tsuneyoshi Kaido and Ohara Kerusauskas Rayel,  are with the CPGEI, Federal University of Technology - Parana, Curitiba, PR, 80230-901, Brazil (e-mail:\{jlrebelatto, richard\}@utfpr.edu.br, rodrigokaido@gmail.com, oharakr@gmail.com).}
\thanks{Bartolomeu~F.~Uchôa-Filho is with the EEL, Federal University of Santa Catarina, Florianópolis, SC, 88040-900, Brazil (e-mail: uchoa@eel.ufsc.br).}%
}%

\begin{document}
\maketitle
\begin{abstract}
We evaluate the secrecy performance of a multiple access cooperative network where the destination node is wiretapped by a malicious and passive eavesdropper. We propose the application of the network coding technique as an alternative to increase the secrecy at the destination node, {\bf on the top of improving the error performance of the legitimate communication, already demonstrated in the literature. Network coding is leveraged}  by assuming that the legitime cooperative nodes are able to perform non-binary linear combinations of different frames before the transmission. Different scenarios with and without channel state information (CSI) at the transmitter side are evaluated.
The effectiveness of the proposed schemes is evaluated in terms of secrecy outage probability through theoretic and numerical analyses. It is shown that, even when the legitimate transmitters do not have any CSI, the secrecy can be increased through the use of network coding when compared to the direct transmission and traditional cooperative techniques.
\end{abstract}
\section{Introduction}
\label{sec:intro}



Information security has become a major concern in wireless communications, due to the broadcast nature of the wireless medium which allows eavesdroppers to potentially intercept any transmission. Information theoretic secrecy, introduced by Shannon in 1949~\cite{shannon.49.secrecy}, is a promising approach towards increasing communication security. In~\cite{wyner.1975}, Wyner elaborated on the work of Shannon by introducing the so-called wiretap channel, which is composed of a pair of legitimate nodes communicating in the presence of an eavesdropper.

Recent works have applied information theoretic secrecy ideas to wireless communications, showing that the randomness inherent to wireless channels can help in improving the secrecy of the network~\cite{gopala.08,barros.06.secrecy,bloch.11,tang.07,tang.09}, under different assumptions regarding channel state information (CSI) at the transmitters. When the transmitters have global CSI regarding the legitimate and eavesdropper links, perfect secrecy is achieved by adapting the rate of the wiretap code~\cite{gopala.08}. When only the CSI of the legitimate channels are available at the transmitter side, as commonly assumed in the literature~\cite{barros.06.secrecy,bloch.11,alves.12.secrecy.tas,pheelep.13,gabry.12,lai.08,kaido.14}, the secrecy needs to be evaluated through a probabilistic analysis, by finding the probability that the information is leaked to the eavesdropper for a given fixed secure transmission rate.  The problem of establishing secure communication in a scenario without any CSI at the transmitter side was also addressed in the literature~\cite{tang.07,tang.09}. In this situation, the secrecy outage probability becomes the union of two independent events: \emph{i.)} the {\it reliability outage event}, when the legitimate receiver could not decode the transmitted message; and \emph{ii.)} the {\it secrecy outage event}, when the instantaneous capacity of the eavesdropper is above the equivocation rate of the considered wiretap code.

Similarly to communication networks without secrecy constraints, the channel conditions dictate the network performance. It is then necessary for the legitimate nodes to have some advantage over the eavesdropper in terms of instantaneous channel quality to guarantee the existence of secure communications. In this regards, many techniques have been recently proposed to increase the secrecy in wireless networks. Some of them consider the use of multiple antennas~\cite{alves.12.secrecy.tas,pheelep.13}, or even adopt the concept of cooperative communications~\cite{gabry.12,lai.08,kaido.14}, which is a technique initially proposed to increase the reliability of wireless communications~\cite{laneman.04,sendonaris.03}. In cooperative networks, the sources help each other by relaying their messages, and the transmission is usually divided in two phases: the so-called {\it broadcast phase} (BP), where the sources broadcast their own information frames (IFs), and the {\it cooperative phase} (CP), where the nodes transmit parity frames (PFs) to the destination, which are composed of redundant information related to their own IFs and/or to the IFs of their partners. One of the most well known cooperative protocols is the decode-and-forward (DF)~\cite{laneman.04}, where the nodes just act as routers in the cooperative phase, relaying the IF from its partner.


In~\cite{lai.08}, the authors presented a pioneering study on the secrecy of cooperative communications, by combining concepts of the relay~\cite{meulen.71} and wiretap~\cite{wyner.1975} channels in the so-called relay-eavesdropper channel, as well as establishing the theoretical bounds for the rate-equivocation of the channel. More recently, the secrecy performance of a cooperative network under the DF protocol was carried out in~\cite{gabry.12}, considering either a passive or an active eavesdropper. It was shown in~\cite{gabry.12} that cooperation is capable of increasing the network secrecy when compared to the direct transmission.


The network coding technique~\cite{ahlswede.00} has been also recently applied to cooperative networks, aiming to increase their reliability~\cite{xiao.10,rebelatto.10.TIT}. In a network-coded cooperative network, the sources transmit linear combinations of different messages instead as just acting as routers. It was shown in~\cite{xiao.10,rebelatto.10.TIT} that, if such linear combinations are performed over a large enough non-binary finite field GF($q$), the system diversity order can be increased when compared to the traditional DF protocol, reducing the system outage probability. 

Motivated by the promising performance of the network coding technique, we presented in~\cite{kaido.14} some preliminary results on the performance of such technique in a scenario subject to secrecy constraints, where two sources aim to cooperatively transmit independent information to a common destination in the presence of a malicious eavesdropper. The results in~\cite{kaido.14}, which assumes CSI of the legitimate channels at the transmitters, indicate the potential of network coding to increase the secrecy.

\subsection{Contributions}

Against the background presented above, the novel contributions of this paper are summarized as follows:
\begin{itemize}
\item  We elaborate on the results from~\cite{kaido.14} by extending in time the code construction, allowing the sources to transmit an arbitrary and independent number of frames either in the broadcast or the cooperative phase. More specifically, we adopt the generalized network coding (GNC) scheme from~\cite{rebelatto.10.TIT} instead of the network coding scheme from~\cite{xiao.10} adopted in~\cite{kaido.14};
\item We also generalize the results in~\cite{kaido.14}  to a network-coded cooperative network with multiple ($M\geq 2$) sources, obtaining closed-form approximations to the secrecy outage probability under the assumption that the sources have CSI of the legitimate channels, but do not know the channel to the eavesdropper, as commonly assumed in the literature;
\item Since even partial CSI is not easy to be obtained in practice, we also calculate the secrecy outage probability of the network-coded cooperative scheme in the situation where the sources do not have any CSI at all, and show that even in this case network coding can be beneficial towards increasing the network secrecy.
\end{itemize}

The rest of this work is organized as follows. In Section~\ref{sec:preliminares} we introduce the system model, while Section~\ref{sec:secrecy} introduces the concept of secrecy outage probability considering the direct non-cooperative communication. Sections~\ref{sec:df} and~\ref{sec:nc} present the DF and GNC cooperative protocols, respectively. The secrecy outage analysis of the GNC scheme is carried out in Section~\ref{sec:secrecy-nc}, which is followed by numerical results in Section~\ref{sec:numerical_results}. Finally, Section \ref{sec:final_comments} concludes the paper.

{\it Notations:} $\log(\cdot)$ denotes base-2 logarithm. $(x)^+$ means $\max\{0,x\}$. Lower-case boldface symbols represent vectors. The symbol $\boxplus$ stands for summation over a finite field $GF($q$)$.

\section{System Model} \label{sec:preliminares}
\label{sec:system_model}
\begin{figure}[!t]
\centering
{
\psfrag{1}[B][B]{S$_1$}
\psfrag{2}[B][B]{S$_2$}
\psfrag{a}[c][c]{{\tiny $h_{21}$}}
\psfrag{b}[c][c]{{\tiny $h_{12}$}}
\psfrag{c}[c][c]{{\tiny $h_{1D}$}}
\psfrag{d}[c][c]{{\tiny $h_{2D}$}}
\psfrag{e}[c][c]{{\tiny $h_{1E}$}}
\psfrag{f}[c][c]{{\tiny $h_{2E}$}}
\psfrag{D}[B][B]{D}
\psfrag{E}[B][B]{E}
\includegraphics[height=3.5cm]{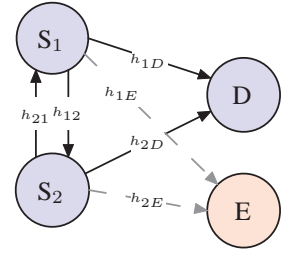}}
\vspace{-0.2cm}
\caption{System model. Multi-source network where nodes (two in the example, referred to as S$_1$ and S$_2$) have independent information to transmit to a common destination (D) in the presence of an eavesdropper (E).}
    \label{fig:system_model_1}
\end{figure}

We consider a multiple access cooperative network composed of $M$ sources having independent information to transmit to a common destination node. We assume the existence of a malicious eavesdropper (also referred to as just E) near the destination node, as illustrated in Fig.~\ref{fig:system_model_1}. Omitting the time index, the signal received by node $j$ after a transmission of signal $\textbf{x}_i$ performed by source $i$ is given by
\begin{equation} \label{eq:system_model}
{\bf y}_j = \sqrt{P_id_{ij}^{-\alpha}}h_{ij} {\bf x}_i + {\bf n}_j,
\end{equation}
where $P_i$ corresponds to the transmission power, $d_{ij}>1$ represents the distance between sources $i$ and $j$, $\alpha$ stands for the path-loss exponent, $h_{ij}$ represents the block-fading coefficient, modeled as a circularly-symmetric complex Gaussian independent and identically distributed random variable (thus $|h_{ij}|$ follows a Rayleigh distribution). The additive white Gaussian noise is represented by ${\bf n}_j$ and we assume unitary bandwidth.

We adopt the notation  $i,j \in \{1,\ldots,M, D, E\}$ when referring to source 1 (S$_1$) to source $M$ (S$_M$), destination (D) and eavesdropper (E), respectively. The instantaneous signal-to-noise ratio (SNR) is defined as
\begin{equation}
\gamma_{ij} = \bar{\gamma}_{ij} |h_{ij}|^2,
\end{equation}
where $\bar{\gamma}_{ij}  =  \frac{P_i}{d_{ij}^{\alpha} \sigma_j^2}$ is the average SNR and $\sigma_j^2$ is the noise variance.  We assume a symmetric scenario in which all source nodes are approximately at the same distance from D; then $\bar{\gamma}_{iD}=\bar{\gamma}_{D} \ \forall \ i \in \{1,2,\ldots,M\}$. Moreover, we also assume that all sources are at approximately the same distance from E, so that $\bar{\gamma}_{iE}=\bar{\gamma}_{E} \ \forall \ i \in \{1,2,\ldots,M\}$.

Without secrecy constraints, assuming Gaussian inputs, an outage event in an individual link occurs when the mutual information $\mathcal{I}~=~\log(1 + \gamma)$ falls below a given target information rate $\mathcal{R}$ (in bits/channel use). The probability of such an event is called {\it outage probability}, and is defined as~\cite{goldsmith.05}
\begin{eqnarray} \label{eq:outage_dt}
\mathcal{O}(\mathcal{R}, \bar{\gamma}) &\triangleq& \Pr\left\{\mathcal{I} < \mathcal{R} \right\} \nonumber \\
&=& \Pr\left\{|h|^2 <\frac{2^\mathcal{R}-1}{\bar{\gamma}}\right\}.
\end{eqnarray}

One can see that the definition of outage probability presented in~\eqref{eq:outage_dt} corresponds to the cumulative density function (CDF) of the random variable $|h|^2$ evaluated at the point $(2^\mathcal{R}-1)/\bar{\gamma}$. As $|h|$ follows a Rayleigh distribution, then $|h|^2$ is exponentially distributed and the outage probability in~\eqref{eq:outage_dt} becomes~\cite{goldsmith.05}
\begin{equation} \label{eq:outage_dt_rayleigh}
\mathcal{O} (\mathcal{R}, \bar{\gamma}) = 1-\exp\left(-\frac{2^\mathcal{R}-1}{\bar{\gamma}}\right).
\end{equation}

Based on the formulation of the individual link outage probability, under the assumption that all the links throughout the network are subject to independent and identically distributed Rayleigh fading, the overall outage probability of a generic scheme $X$ can be approximated for high SNR region as~\cite{rebelatto.10.TIT}
\begin{equation} \label{eq:outage_generic}
\begin{split}
\tilde{\mathcal{O}}_{X}(\mathcal{R}, \bar{\gamma}) & = \mu_X \big[\mathcal{O} (\mathcal{R}, \bar{\gamma})\big]^{\mathcal{D}_X}\\
& = \mu_X \left[1-\exp\left(-\frac{2^\mathcal{R}-1}{\bar{\gamma}}\right) \right]^{\mathcal{D}_X},
\end{split}
\end{equation}
where $\mu_X$ and $\mathcal{D}_X$ correspond to the \textit{coding gain} and \textit{diversity order} of scheme $X$, respectively. Moreover, the diversity order $\mathcal{D}_X$ is formally defined as~\cite{goldsmith.05}
\begin{equation} \label{eq:diversity_oder}
\mathcal{D}_X \triangleq \lim_{\bar{\gamma}\rightarrow \infty} \frac{-\log \mathcal{O}_{X}(\mathcal{R}, \bar{\gamma})}{\log \bar{\gamma}}.
\end{equation}

In order to perform a fair comparison between different protocols, one must take into account the multiplexing loss inherent to many cooperative schemes~\cite{laneman.04}. Thus, we consider that the target information rate of the generic cooperative  protocol $X$ is given by $\mathcal{R}_X \triangleq \mathcal{R}/R_X$, where ${\cal R}$ is the attempted transmission rate in the case of non-cooperative direct transmission, and $R_X$ corresponds to the code rate of the protocol $X$, defined as the ratio between the number of time slots allocated to the transmission of new data and the total number of time slots used by the protocol, with $0 \leq R_X \leq 1$. For the direct transmission, $R_X=R_{\text{DT}} = 1$.


\section{Secrecy Outage Probability (SOP)}\label{sec:secrecy}

In the case where the sources have global channel state information (CSI), as in~\cite{gopala.08}, perfect secrecy can be achieved by adapting the rate of the wiretap code according to the instantaneous channels realization (seen at both the legitimate and the eavesdropper). However, since assuming the knowledge of instantaneous channel condition of the eavesdropper might not be practical in several cases, in what follows we evaluate two more realistic scenarios regarding the availability of CSI at the source nodes: {\it i)} The sources have CSI of the legitimate channels only; {\it ii)} The sources have no CSI at all.

\subsection{Sources with partial CSI}
When only the CSI of the legitimate links is available at the sources, perfect secrecy cannot be guaranteed since the instantaneous channel information from E is unknown. Thus, a probabilistic secrecy analysis must be carried out~\cite{barros.06.secrecy}, by determining the probability that E successfully spies some amount of information, for a given fixed secure transmission rate (which implies a variable transmission rate over the legitimate channel following its instantaneous capacity). It is noteworthy that this is the scenario assumed in most papers addressing secrecy outage probability in the literature, as for instance~\cite{barros.06.secrecy,alves.12.secrecy.tas,pheelep.13}.

More specifically, following the wiretap code construction described for instance in \cite{bloch.11}, if we assume that the channel condition of the legitimate destination is known at the sources, we can design a wiretap code with block length $n$ that contains $2^{n{\cal R}}$ codewords, where the transmission rate ${\cal R}$ is made equal to ${\cal C_D}$, the instantaneous channel capacity seen at the legitimate destination. Moreover, we set a number of codewords per bin in the wiretap code equal to $2^{n{\cal R}_E}$, where ${\cal R}_E$ is the eavesdropper's equivocation rate. The rate of secure communication is then ${\cal R}_s = {\cal R} - {\cal R}_E = {\cal C}_D - {\cal R}_E$, which is usually fixed, implying that ${\cal R}_E = {\cal C}_D - {\cal R}_s$ varies according to the channel condition seen at the legitimate destination.

Therefore, a secrecy outage event occurs when the instantaneous eavesdropper's channel capacity, ${\cal C}_E$, exceeds the equivocation rate ${\cal R}_E$, or alternatively when the difference between the instantaneous capacities of the main and the eavesdropper channels ${\cal C}_D-{\cal C}_E$ is lower than ${\cal R}_s$ \cite{bloch.11}. Considering Gaussian inputs, the instantaneous channel capacities are equal to the instantaneous mutual information seen at the destination and at the eavesdropper, respectively $\mathcal{I}_{D}$ and $\mathcal{I}_{E}$, so that the instantaneous secrecy capacity is finally written as~\cite{barros.06.secrecy,bloch.11}
\begin{equation}\label{eq:secrecy_capacity}
\mathcal{C}_{s} = \left(\mathcal{I}_{D} - \mathcal{I}_{E}\right)^{+}.
\end{equation}
From~\eqref{eq:secrecy_capacity}, it follows that the instantaneous  secrecy capacity is positive when $\gamma_{D} > \gamma_{E}$ and equal to zero when $\gamma_{D}~\leq~\gamma_{E}$. Thus, the probability of existence of a non-zero secrecy capacity is
\begin{eqnarray} \label{eq:cs_0}
\Pr\{\mathcal{C}_s > 0\} &=& \Pr\{\gamma_{D} > \gamma_{E}\}
\\
&=& \int_0^{\infty} \int_0^{\gamma_{D}} p_{\gamma_{D}\gamma_{E}}(\gamma_{D},\gamma_{E}) d\gamma_{E}d\gamma_{D} \nonumber \\
&=& \int_0^{\infty} \int_0^{\gamma_{D}} p_{\gamma_{D}}(\gamma_{D})p_{\gamma_{E}}(\gamma_{E}) d\gamma_{E}d\gamma_{D}. \nonumber
\end{eqnarray}

Moreover, the {\it secrecy outage probability} (SOP) is the probability that $\mathcal{C}_{s}$ is less than a target secrecy rate $\mathcal{R}_s$~\cite{bloch.11}:
\begin{equation} \label{eq:cs_rs}
\mathcal{S} = \Pr\left\{\mathcal{C}_s < \mathcal{R}_s \right\}.
\end{equation}

For Rayleigh fading in the non-cooperative direct transmission (DT), the pdf and CDF of the random variable $\gamma_D$ (the same holds to $\gamma_E$) are given by~\cite{goldsmith.05}
\begin{subequations}
\begin{equation} \label{eq:cdf_dt}
F_{\gamma_{D}}(\gamma_{D}) = 1-\exp\left(-\frac{\gamma_{D}}{\bar{\gamma}_D}\right),
\end{equation}
\begin{eqnarray} \label{eq:pdf_dt}
p_{\gamma_{D}}(\gamma_{D}) &=& \frac{\partial \left[ F_{\gamma_{D}}(\gamma_{D})\right]}{\partial \gamma_{D}} \nonumber \\
&=& \frac{1}{\bar{\gamma}_D} \exp\left(-\frac{\gamma_{D}}{\bar{\gamma}_D}\right),
\end{eqnarray}
\end{subequations}
leading to the following probability of existence of non-zero secrecy capacity~\cite{barros.06.secrecy}
\begin{equation}\label{eq:cs_0_dt}
\Pr\{\mathcal{C}_{s,\text{DT}} > 0\} = \frac{\bar{\gamma}_D}{\bar{\gamma}_D + \bar{\gamma}_E}.
\end{equation}

The SOP, in turn, was shown in~\cite{barros.06.secrecy} to be
\begin{equation} \label{eq:cs_rs_dt}
\mathcal{S}_{\text{DT}}^{\text{csi}} = 1-\frac{\bar{\gamma}_{D}}{\bar{\gamma}_{D}+2^{\mathcal{R}}\bar{\gamma}_{E}}
\exp\left(- \frac{2^{\mathcal{R}}-1}{\bar{\gamma}_{D}} \right).
\end{equation}

\subsection{Sources without CSI}

In this case we assume that the transmitters do not have CSI of either the legitimate or the eavesdropper channels. Thus, the instantaneous capacity seen at the legitimate destination, ${\cal C}_D$, is not known at the transmitters and therefore we must choose a fixed total number of $2^{n{\cal R}}$ codewords and a fixed number of codewords per bin  equal to $2^{n{\cal R}_E}$ in the wiretap code. The fixed rate of attempted secure communication is then ${\cal R}_s = {\cal R} - {\cal R}_E$. Therefore, an outage corresponds to the occurrence of any of the following two independent events~\cite{tang.07}: {\it i)} The instantaneous channel capacity seen at the legitimate destination is smaller than ${\cal R}$. Such an event is referred to as {\it reliability outage} and has probability $\Pr\{\mathcal{I}_{D}\!<\!\mathcal{R}\}$; {\it ii)} The instantaneous channel capacity seen at eavesdropper is larger than the equivocation rate $\mathcal{R}_E$ of the wiretap code, so that the eavesdropper is able to recover at least part of the information intended to the legitimate transmitter. This event is referred to as {\it secrecy outage}, and has probability of occurrence given by $\Pr\{\mathcal{I}_{E}\!\geq\! \mathcal{R}_E\}$.


The overall secrecy outage probability is then given by the union of the two aforementioned independent  events, being represented by
\begin{equation} \label{eq:ps-nocsi1}
\begin{split}
\mathcal{S}^{\text{no-csi}} &= \Pr\left\{(\mathcal{I}_{D}<\mathcal{R})\bigcup (\mathcal{I}_{E}\geq \mathcal{R}_E) \right\} \\
&= \Pr\left\{\mathcal{I}_{D}<\mathcal{R}\right\}+ \Pr\left\{\mathcal{I}_{E}\geq \mathcal{R}_E \right\} \\
&\qquad - \Pr\{\mathcal{I}_{D}<\mathcal{R}, \mathcal{I}_{E}\geq \mathcal{R}_E\}.
\end{split}
\end{equation}

Generically, the secrecy outage probability of a given scheme $X$ is obtained according to~\eqref{eq:ps-nocsi1} as
\begin{equation} \label{eq:ps-nocsi2}
\begin{split}
\mathcal{S}_{X}^{\text{no-csi}} &= \mathcal{O}_{X}(\mathcal{R}, \bar{\gamma}_D) + \left[ 1-\mathcal{O}_{X}(\mathcal{R}_E, \bar{\gamma}_E)\right] \\
& \qquad - \mathcal{O}_{X}(\mathcal{R}, \bar{\gamma}_D)\left[ 1-\mathcal{O}_{X}(\mathcal{R}_E, \bar{\gamma}_E)\right],
\end{split}
\end{equation}
where $\mathcal{O}_{X}(\mathcal{R}, \bar{\gamma})$ is the reliability outage probability of scheme~$X$.

For the direct transmission, through the substitution of~\eqref{eq:outage_dt_rayleigh} in~\eqref{eq:ps-nocsi2}, one can show that the SOP when the sources do not have any CSI is
\begin{equation} \label{eq:ps-nocsi-dt}
\mathcal{S}_{\text{DT}}^{\text{no-csi}} = 1 \!-\! \exp\left(\!-\frac{2^{\mathcal{R}}-1}{\bar{\gamma}_D}\right) \! \left[1 \!-\! \exp\left(\!-\frac{2^{\mathcal{R}_E}-1}{\bar{\gamma}_E}\right)\!\right].
\end{equation}


\section{Decode-and-Forward (DF)} \label{sec:df}

\begin{figure*}[!t]
\centering
{\scriptsize
\subfigure[Decode-and-Forward (DF)\label{fig:time_df}]{
\psfrag{1}[B][b]{ S$_1$}
\psfrag{2}[B][b]{ S$_2$}
\psfrag{t}[B][B]{{ time}}
\psfrag{a}[B][B]{ $T/2$}
\psfrag{p}[B][c]{{ \it Broadcast Phase}}
\psfrag{q}[B][c]{{ \it Cooperative Phase}}
\psfrag{d}[B][B]{$I_1$}
\psfrag{e}[B][B]{$I_2$}
\psfrag{f}[B][B]{$I_2$}
\psfrag{g}[B][B]{$I_1$}
\includegraphics[width=0.49\textwidth]{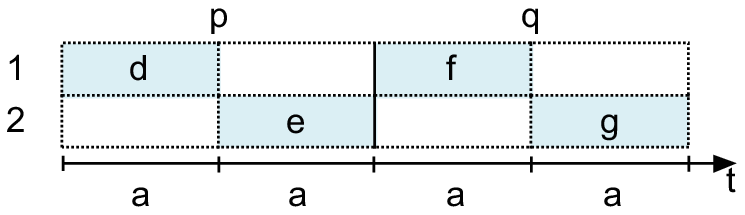}}
\subfigure[Network Coding (NC)\label{fig:time_nc}]{
\psfrag{1}[B][b]{ S$_1$}
\psfrag{2}[B][b]{ S$_2$}
\psfrag{t}[B][B]{{ time}}
\psfrag{a}[B][B]{$T/2$}
\psfrag{p}[B][c]{{ \it Broadcast Phase}}
\psfrag{q}[B][c]{{ \it Cooperative Phase}}
\psfrag{d}[B][B]{$I_1$}
\psfrag{e}[B][B]{$I_2$}
\psfrag{f}[B][B]{$I_1\boxplus I_2$}
\psfrag{g}[B][B]{$I_1\boxplus2I_2$}
\includegraphics[width=0.49\textwidth]{fig/schemesTime.eps}}
\subfigure[Generalized Network Coding (GNC)\label{fig:time_gnc}]{
\psfrag{D}[B][b]{ D}
\psfrag{1}[B][b]{ S$_1$}
\psfrag{2}[B][b]{ S$_2$}
\psfrag{t}[B][B]{{ time}}
\psfrag{a}[B][B]{ $T/2$}
\psfrag{p}[B][c]{{ \it Broadcast Phase $1$}}
\psfrag{q}[B][c]{{ \it Broadcast Phase $k_1$}}
\psfrag{r}[B][c]{{ \it Cooperative Phase $1$}}
\psfrag{s}[B][c]{{ \it Cooperative Phase $k_2$}}
\psfrag{d}[B][B]{$I_1[1]$}
\psfrag{e}[B][B]{$I_2[1]$}
\psfrag{f}[B][B]{$I_1[k_1]$}
\psfrag{g}[B][B]{$I_2[k_1]$}
\psfrag{x}[B][B]{$\boxplus_1[k_1\!\!+\!\!1]$}
\psfrag{y}[B][B]{$\boxplus_2[k_1\!\!+\!\!1]$}
\psfrag{z}[B][B]{$\boxplus_1[k_1\!\!+\!\!k_2]$}
\psfrag{w}[B][B]{$\boxplus_2[k_1\!\!+\!\!k_2]$}
\includegraphics[width=0.90\textwidth]{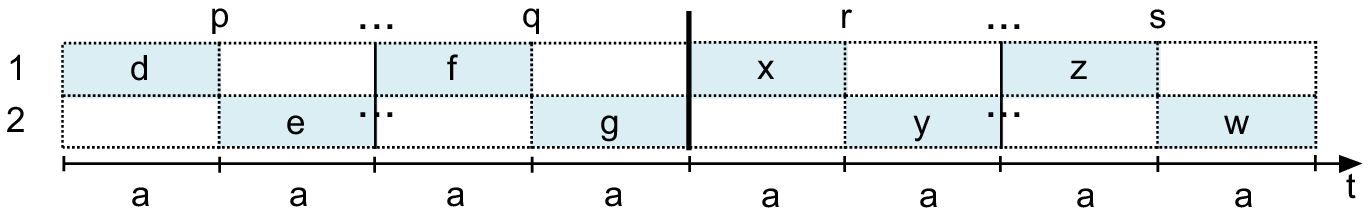}}
}
\vspace{-0.2cm}
\caption{Two-source time division channel allocation considering (a) Decode-and-Forward (DF) protocol; (b) Network Coding-based (NC) protocol; (c) Generalized Network Coding-based (GNC) protocol. $T$ represents the time-slot duration and the symbol $\boxplus$ in (b) stands for summation over a finite field. In (c), $\boxplus_i[k]$ corresponds to a linear combination transmitted by source $i$ at time slot $k$, which is composed of all the IFs received during the broadcast phase, including source's $i$ own IFs.}
\label{fig:time-division}
\end{figure*}

The decode-and-forward (DF) is a well established cooperative protocol, being largely investigated in recent works on cooperative communication, {\it e.g.}~\cite{laneman.04,sendonaris.03,gabry.12,lai.08}. In the DF protocol, after broadcasting their own IFs in the first time slot, each source retransmits a given  partner's IF in the cooperative phase, usually adopting the same codebook (repetition code) as the source, as illustrated in Fig.~\ref{fig:time_df} for a two-source network. Upon receiving two copies of the same message, the receiver performs maximum ratio combining (MRC) to optimally combine both observations. One can see that the code rate of DF is given by $R_{\text{DF}} = 1/2$, since one PF is transmitted in the cooperative phase for every IF broadcasted previously. Thus, in order to perform a fair comparison to the DT scheme, one must transmit with a transmission rate twice that of DT.

\subsection{Sources with partial CSI}
When the sources have CSI of the legitime channels, the instantaneous secrecy capacity of the DF scheme is~\cite{gabry.12}
\begin{equation} \label{eq:cs_df}
\mathcal{C}_{s,\text{DF}} =\frac{1}{2}\Big(\log_2(1+ \gamma_{D}) - \log_2(1+ \gamma_{E}) \Big)^{+},
\end{equation}
where $\gamma_{D}$ and $\gamma_{E}$ are respectively the instantaneous SNR at D and E after combining the messages. Assuming that the channels between the sources are error-free\footnote{Note that this is an optimistic assumption that favors the DF scheme when compared to the other schemes. As our goal is to show that the network coding technique can outperform traditional cooperation protocols as DF, this assumption does not invalidate our results.} and focusing on the message of S$_1$ (the same result is valid to the other source due to the symmetry), the instantaneous SNR seen at D and E after performing MRC is~\cite{goldsmith.05}
\begin{subequations}
\begin{equation}\label{eq:gamma_d_df}
\begin{split}
\gamma_{D} &=
    \bar{\gamma}_{1D} |h_{1D}|^2 + \bar{\gamma}_{2D} |h_{2D}'|^2, \\
    &=
    \bar{\gamma}_{D}\left( |h_{1D}|^2 + |h_{2D}'|^2\right),
\end{split}
\end{equation}
\begin{equation}\label{eq:gamma_e_df}
\begin{split}
\gamma_{E} &=
    \bar{\gamma}_{1E} |h_{1E}|^2 + \bar{\gamma}_{2E} |h_{2E}'|^2, \\
    &=
    \bar{\gamma}_{E} \left(|h_{1E}|^2 + |h_{2E}'|^2\right),
\end{split}
\end{equation}
\end{subequations}
where the superscript $'$ refers to the channel realization in the cooperative phase ($h_{iD}$ and $h_{iD}'$ are assumed to be independent). From the results in~\cite{laneman.04,barros.06.secrecy}, it can be shown that the SOP of the two-source DF scheme in this case is
\begin{equation} \label{eq:outage_df1}
\begin{split}
\mathcal{S}_{\text{DF}}^{\text{csi}} &= \Pr\{\mathcal{C}_{s,\text{DF}} < \mathcal{R}_s\}  \\
&= 1 - \frac{\bar{\gamma}_{D}}{ \left(\bar{\gamma}_{D} + \xi\,\bar{\gamma}_{E}\right)^{3}}\, \exp\left(-\frac{\xi-1}{\bar{\gamma}_{D}}\right)  \\
&\quad \times  \Big[\bar{\gamma}_{D}\,(\xi - 1 + \bar{\gamma}_{D}) + \xi\,\bar{\gamma}_{E}\,(\xi\!-1\! +\! 3\,\,\bar{\gamma}_{D})\Big],
\end{split}
\end{equation}
where $\xi = 2^{2\mathcal{R}_s}$.

\subsection{Sources without CSI}

When the sources do not have any CSI, the exact SOP of the two-source DF scheme is obtained from~\eqref{eq:ps-nocsi2} as
\begin{equation} \label{eq:ps-nocsi-df}
\begin{split}
\mathcal{S}_{\text{DF}}^{\text{no-csi}} &= \mathcal{O}_{\text{MRC}}(2\mathcal{R}, \bar{\gamma}_D) + \left[ 1-\mathcal{O}_{\text{MRC}}(2\mathcal{R}_E, \bar{\gamma}_E)\right] \\
& \qquad - \mathcal{O}_{\text{MRC}}(2\mathcal{R}, \bar{\gamma}_D)\left[ 1-\mathcal{O}_{\text{MRC}}(2\mathcal{R}_E, \bar{\gamma}_E)\right]\\
& = 1- \left[1\!-\!\exp\left(\!-\frac{2^{2\mathcal{R}_E}-1}{\bar{\gamma}_E}\!\right)\left[1\!+\!\frac{2^{2\mathcal{R}_E}-1}{\bar{\gamma}_E}\right]\right]\\
& \times \exp\left(\!-\frac{2^{2\mathcal{R}}-1}{\bar{\gamma}_D}\!\right)\left[1\!+\!\frac{2^{2\mathcal{R}}-1}{\bar{\gamma}_D}\right],
\end{split}
\end{equation}
where $\mathcal{O}_{\text{MRC}}(\mathcal{R},\bar{\gamma})$ corresponds to the 2-branch outage probability of the MRC scheme, which is given by~\cite{goldsmith.05}
\begin{equation} \label{eq:outage_mrc}
\mathcal{O}_{\text{MRC}}(\mathcal{R},\bar{\gamma}) = 1\!-\!\exp\left(\!-\frac{2^{\mathcal{R}}-1}{\bar{\gamma}}\!\right)\left[1\!+\!\frac{2^{\mathcal{R}}-1}{\bar{\gamma}}\right].
\end{equation}


%
%

\section{Network-Coded Cooperation (NC)}\label{sec:nc}

In a non-binary network-coded (NC) based cooperative protocol with two sources, instead of just acting as routers, the sources are able to transmit linear combinations of all the available IFs during the cooperative phase, as illustrated in Fig.~\ref{fig:time_nc}. If such linear combinations are performed over a high enough finite field, it is shown in~\cite{xiao.10} that gains in terms of diversity order can be achieved over the DF scheme.

Let us focus again on the message from S$_1$ and focus for the moment being in the two-source case. If the intersource channel is not in outage (which happens with probability $1-\mathcal{O}(2\mathcal{R}, \bar{\gamma}_D)$, where $1-\mathcal{O}(2\mathcal{R}, \bar{\gamma}_D)$ is the outage probability of an individual link obtained from~\eqref{eq:outage_dt_rayleigh}), we can see that D is able to recover S$_1$'s message from any two out the following four received frames: $I_1, \; I_2, \; I_1 \boxplus I_2,\; I_1\boxplus2I_2$ (the symbol $\boxplus$ stands for summation over a finite field). The information frame from S$_1$ is not recovered by D when the direct transmission and at least two out of the three remaining packets cannot be decoded, which happens with probability~\cite{xiao.10}
\begin{equation}
\mathcal{O}_1(\mathcal{R}, \bar{\gamma}_D)) \approx 3\big[\mathcal{O}(2\mathcal{R}, \bar{\gamma}_D)\big]^3.
\end{equation}

When the channel between S$_1$ and S$_2$ is in outage (which happens with probability $\mathcal{O}(2\mathcal{R}, \bar{\gamma}_D)$), S$_1$ and S$_2$ retransmit their own messages in the cooperative phase. Upon receiving two copies of the same message, we assume that D performs MRC, leading to the outage probability presented in~\eqref{eq:outage_mrc}.

The overall outage probability of the NC scheme was shown in~\cite{xiao.10} to be
\begin{equation} \label{eq:outage_nc1}
\begin{split}
\mathcal{O}_{\text{NC}}(\mathcal{R}, \bar{\gamma}_D) &= \left[1-\mathcal{O}(2\mathcal{R}, \bar{\gamma}_D)\right]\, \mathcal{O}_1(\mathcal{R}, \bar{\gamma}_D) \\
 & \qquad + \quad \mathcal{O}(2\mathcal{R}, \bar{\gamma}_D)\, \mathcal{O}_{\text{MRC}}(2\mathcal{R}, \bar{\gamma}_D)\\
&  \approx  3.5\left[1-\exp\left(-\frac{2^{2\mathcal{R}}-1}{\bar{\gamma}_{D}}\right) \right]^3,
\end{split}
\end{equation}
where the approximation holds for the high SNR region. We can see from~\eqref{eq:outage_nc1} that diversity order of 3 is achieved, in contrast to the diversity order of 2 obtained by the DF scheme~\cite{xiao.10}.

In~\cite{rebelatto.10.TIT}, a generalization of the scheme proposed in~\cite{xiao.10} was presented, as illustrated in Fig.~\ref{fig:time_gnc}. In the scheme proposed in~\cite{rebelatto.10.TIT}, referred to as generalized network coding (GNC), the sources are able to broadcast $k_1$ IFs in the broadcast phase, as well as transmit an arbitrary number $k_2$ of  PFs in the cooperative phase, leading to a more flexible network code rate given by~\cite{rebelatto.10.TIT}
\begin{equation} \label{eq:rate_gnc}
R_{\text{GNC}} = \frac{k_1}{k_1+k_2}.
\end{equation}

 In this scenario with two sources, when the intersource channel is not in outage (which happens with probability $1-\mathcal{O}(\mathcal{R}/R_{\text{GNC}}, \bar{\gamma}_D)$), the destination receives $2(k_1\!+\!k_2)$ frames (each source broadcasts $k_1$ IFs and then transmits $k_2$ PFs in the cooperative phase), and a given IF is not recovered by D when the direct transmission and at least $2k_2$ out of the remaining $2(k_2\!+\!k_1)\!-\!1$ frames cannot be decoded, which happens with probability~\cite{rebelatto.10.TIT}
\begin{equation} \label{eq:outage_gnc_exact_p1}
\mathcal{O}_{1}(\mathcal{R}, \bar{\gamma}_D) = \mathcal{O} \sum_{i=0}^{2k_1-1} \mu_{\text{GNC}_1}(i) \, \mathcal{O}^{2k_2+i}[1\!-\!\mathcal{O}]^{2k_1-1-i},
\end{equation}
where $\mathcal{O}$ is the short to $\mathcal{O}(\mathcal{R}/R_{\text{GNC}}, \bar{\gamma}_D)$ and $\mu_{\text{GNC}_1}(i) = {2k_2\!+\!2k_1\!-\!1 \choose 2k_2+i}$ corresponds to a binomial coefficient that takes into account the multiplicity of independent events that lead to the same outage probability.

When the intersource channel is in outage (which happens with probability $\mathcal{O}_{12} = \mathcal{O}(\mathcal{R}/R_{\text{GNC}}, \bar{\gamma}_D)$), a given IF is not recovered by D when the direct transmission and at least $k_2$ out of the $k_2\!+\!k_1\!-1$ remaining frames cannot be decoded, which happens with probability~\cite{rebelatto.10.TIT}
\begin{equation} \label{eq:outage_gnc_exact_p2}
\mathcal{O}_{2}(\mathcal{R}, \bar{\gamma}_D) = \mathcal{O} \sum_{i=0}^{k_1-1} \mu_{\text{GNC}_2}(i)\, \mathcal{O}^{k_2+i}[1\!-\!\mathcal{O}]^{k_1-1-i},
\end{equation}
where $\mu_{\text{GNC}_2}(i) = {k_2\!+\!k_1\!-\!1 \choose k_2+i}$. The overall outage probability of the two-source GNC scheme is then given by:
\begin{equation}\label{eq:outage_gnc_exact}
\begin{split}
&\mathcal{O}_{\text{GNC}}(\mathcal{R}, \bar{\gamma}_D)=\\
 &= (1\!-\!\mathcal{O}_{12}) \,\mathcal{O}_{1}(\mathcal{R}, \bar{\gamma}_D) \!+\! \mathcal{O}_{12} \, \mathcal{O}_{2}(\mathcal{R}, \bar{\gamma}_D)\\
&= (1\!-\!\mathcal{O}_{12})\mathcal{O}\!\!\sum_{i=0}^{2k_1-1} \mu_1(i) \, \mathcal{O}^{2k_2+i}[1\!-\!\mathcal{O}]^{2k_1-1-i}\\
& \quad + \mathcal{O}_{12}\mathcal{O} \sum_{i=0}^{k_1-1} \mu_2(i) \, \mathcal{O}^{k_2+i}[1\!-\!\mathcal{O}]^{k_1-1-i}.
\end{split}
\end{equation}

\subsection{Extension to the $M$-source scenario}

 When the intersource channels are assumed to be outage-free, the outage probability of the $M$-source GNC scheme is obtained through the generalization of~\eqref{eq:outage_gnc_exact_p1} as
 \begin{equation} \label{eq:outage_gnc_exact_free}
\mathcal{O}_{1}(\mathcal{R}, \bar{\gamma}_D) = \mathcal{O} \sum_{i=0}^{Mk_1-1} \mu_{\text{GNC}_1}(i) \, \mathcal{O}^{Mk_2+i}[1\!-\!\mathcal{O}]^{Mk_1-1-i},
\end{equation}
where $\mu_{\text{GNC}_1}(i) = {Mk_2\!+\!Mk_1\!-\!1 \choose Mk_2 + i}$.  As presented in~\cite{rebelatto.10.TIT}, the outage probability from~\eqref{eq:outage_gnc_exact_free} can be accurately approximated for the large SNR region as:
 \begin{equation} \label{eq:outage_gnc_free}
\mathcal{O}_{\text{GNC}}(\mathcal{R}, \bar{\gamma}_D) \approx \mu_{\text{GNC}_1} \left[1-\exp\left(-\frac{2^{\mathcal{R}/R_{\text{GNC}}}-1}{\bar{\gamma}_{D}}\right)\right]^{Mk_2+1},
 \end{equation}
where $\mu_{\text{GNC}_1} = {Mk_2\!+\!Mk_1\!-\!1 \choose Mk_2}$.  Moreover, according to~\cite{rebelatto.10.TIT}, when the intersource channels are subject to outages, the diversity is reduced and the approximation for the outage probability of a network with $M$ sources operating under the GNC scheme with parameters $(k_1,k_2)$ becomes
\begin{equation} \label{eq:outage_gnc}
\tilde{\mathcal{O}}_{\text{GNC}}(\mathcal{R}, \bar{\gamma}_D) \approx \mu_{\text{GNC}_2} \left[1-\exp\left(-\frac{2^{\mathcal{R}/R_{\text{GNC}}}-1}{\bar{\gamma}_{D}}\right)\right]^{M+k_2},
\end{equation}
where $\mu_{\text{GNC}_2} = {k_1 + k_2 - 1 \choose k_2}$. From~\eqref{eq:outage_gnc}, it can be seen that the diversity order of the GNC scheme is $M+k_2$. By properly choosing the values of $k_1$ and $k_2$, it can be shown that the GNC scheme can achieve both code rate and diversity order larger than the NC scheme. It is also worthy mentioning that the GNC scheme reduces to the NC scheme when $k_1=k_2=1$.

Regarding the code design, the maximum diversity order of the GNC scheme is guaranteed if the coefficients of the linear combinations are chosen from a maximum distance separable (MDS) code\footnote{Note that the MDS code is applied on the top of the Wiretap coding applied in the physical layer.}~\cite{rebelatto.10.TIT}.

\section{Network-Coded Cooperation with Secrecy Constraints} \label{sec:secrecy-nc}

In what follows we evaluate the SOP of the GNC scheme considering both the situations with partial CSI and without any CSI at the legitimate sources.

\subsection{Sources with partial CSI}

In this situation, the sources know the legitimate channels and the SOP for the GNC scheme is obtained according to~\eqref{eq:cs_rs} as
\begin{equation} \label{eq:cs_rs_nc_def}
\begin{split}
\mathcal{S}_{\text{GNC}} &= \Pr\left\{\mathcal{C}_{s,\text{GNC}} < \mathcal{R}_s \right\}  \\
 &= \Pr\left\{R_{\text{GNC}}\Big(\log_2(1+ \gamma_{D}) - \log_2(1+ \gamma_{E}) \Big)^{+} < \mathcal{R}_s \right\}  \\
 &= \Pr\left\{\gamma_{D} < 2^{\mathcal{R}_s/R_{\text{GNC}}}(1+\gamma_{E})-1 \right\}  \\
  &= \int_0^{\infty} \int_0^{\gamma_{U}} p_{\gamma_{D}, \gamma_{E}}(\gamma_{D}, \gamma_{E}) d\gamma_{D} d\gamma_{E} \\
  &= \int_0^{\infty} \int_0^{\gamma_{U}} p_{\gamma_{D}}(\gamma_{D})p_{\gamma_{E}}(\gamma_{E}) d\gamma_{D} d\gamma_{E} \\
  &= \int_0^{\infty} F_{\gamma_{D}}(\gamma_U)p_{\gamma_{E}}(\gamma_{E}) d\gamma_{E},
  \end{split}
\end{equation}
where $\gamma_U=2^{\mathcal{R}_s/R_{\text{GNC}}}(1+\gamma_{E})-1$.

According to~\eqref{eq:cs_rs_nc_def}, in order to calculate the SOP, one must have the CDF of $\gamma_{D}$ and the pdf of $\gamma_{E}$. The CDF is directly obtained from the outage probability, whose exact value for a two-source GNC scheme is presented in~\eqref{eq:outage_gnc_exact}, leading to
\begin{equation}
F_{\gamma_{D}}(\gamma_{D}) = \mathcal{O}_{\text{GNC}}(\mathcal{R}, \bar{\gamma}_D).
\end{equation}

The pdf of $\gamma_{E}$, in turn,  can be obtained by differentiating the outage probability (CDF) at the eavesdropper from~\eqref{eq:outage_gnc_exact}:
\begin{equation}
p_{\gamma_{E}}(\gamma_{E}) = \frac{\partial \left[\mathcal{O}_{\text{GNC}}(\mathcal{R}_E, \bar{\gamma}_E)\right]}{\partial \gamma_{E}}.
\end{equation}

Note that the exact outage probability in \eqref{eq:outage_gnc_exact} is composed of several terms which results in a integration with multiple parts in \eqref{eq:cs_rs_nc_def} and finally in a long expression which leads to little insights. Besides that,~\eqref{eq:outage_gnc_exact} is restricted to the two-source case so that obtaining a generic closed-form exact expression is a tough task. Thus, in order to ease the analysis and the comprehension of the final results, in what follows we resort to (as will be shown shortly) a tight approximation through the following assumptions: {\it i)} We consider that the legitimate network is subject to outages in the intersource channels, so that the CDF of $\gamma_{D}$ is approximated by~\eqref{eq:outage_gnc}; {\it ii)} The CDF of E is approximated by~\eqref{eq:outage_gnc_free}, which is an assumption that favors E, and makes the analysis more tractable, since the outages in the intersource channel do not depend on the rate $\mathcal{R}_E$, otherwise one would have to distinguish between the outages in the intersource channels and the outages in the direct channels when calculating the overall outage at E.

The CDF and pdf of $\gamma_{D}$ are then approximated by\footnote{Note that $\mu_{\text{GNC}_1}$ and $\mu_{\text{GNC}_2}$ are made equal to one in \eqref{eq:cdf-pdf-d} and \eqref{eq:cdf-pdf-e} in order to limit the CDF and pdf to the unity and to have unity area, respectively, because the expressions in~\eqref{eq:outage_gnc} and~\eqref{eq:outage_gnc_free} are approximations for the high SNR and are not naturally limited to such values.}
\begin{subequations} \label{eq:cdf-pdf-d}
\begin{equation} \label{eq:cdf_d}
F_{\gamma_{D}}(\gamma_{D}) \approx
 \left[1-\exp\left(-\frac{\gamma_{D}}{\bar{\gamma}_{D}}\right)\right]^{M+k_2}.
\end{equation}
\begin{equation} \label{eq:pdf_d}
\begin{split}
p_{\gamma_{D}}(\gamma_{D}) &= \frac{\partial \left[ F_{\gamma_{D}}(\gamma_{D})\right]}{\partial \gamma_{D}} \\
& \approx \frac{M\!+\!k_2 }{\bar{\gamma}_{D}} \exp\left(-\frac{\gamma_{D}}{\bar{\gamma}_{D}}\right)\\ & \qquad \times\left[1-\exp\left(-\frac{\gamma_{D}}{\bar{\gamma}_{D}}\right)\right]^{M+k_2-1}.
\end{split}
\end{equation}
\end{subequations}
The CDF and pdf of $\gamma_{E}$, in turn, are considered to be
\begin{subequations}\label{eq:cdf-pdf-e}
\begin{equation} \label{eq:cdf_e}
F_{\gamma_{E}}(\gamma_{E}) \approx
 \left[1-\exp\left(-\frac{\gamma_{E}}{\bar{\gamma}_{E}}\right)\right]^{Mk_2+1}.
\end{equation}
\begin{equation} \label{eq:pdf_e}
\begin{split}
p_{\gamma_{E}}(\gamma_{E}) &= \frac{\partial \left[ F_{\gamma_{E}}(\gamma_{E})\right]}{\partial \gamma_{E}} \\
& \approx \frac{Mk_2\!+\!1  }{\bar{\gamma}_{E}} \exp\left(-\frac{\gamma_{E}}{\bar{\gamma}_{E}}\right)\\ & \qquad \times\left[1-\exp\left(-\frac{\gamma_{E}}{\bar{\gamma}_{E}}\right)\right]^{Mk_2}.
\end{split}
\end{equation}
\end{subequations}

By replacing the CDF and pdf of both $\gamma_{D}$ and $\gamma_{E}$ in~\eqref{eq:cs_rs_nc_def}, one can obtain an approximate expression for the SOP of the GNC scheme as presented in the following theorem.
\begin{theorem} \label{th:ps-csi-gnc}
The SOP of the GNC scheme with partial CSI is approximated as
\begin{equation} \label{eq:ps-csi-gnc}
\begin{split}
\mathcal{S}_{\text{\emph{GNC}}}^{\text{csi}}   &\approx (Mk_2\!+\!1)\!\!\sum_{i=0}^{M\!+\!k_2}\! {M\!+\!k_2 \!\choose\! i} [-1]^i \exp\!\left(\!-\frac{\xi-1}{\bar{\gamma}_{D}}i\right)  \\
& \quad \times \quad \textsc{B}\left(\frac{\xi\,\bar{\gamma}_{E}i + \bar{\gamma}_{D} }{\bar{\gamma}_{D}},Mk_2\!+\!1\right),
\end{split}
\end{equation}
where $\xi=2^{\mathcal{R}_s/R_{\text{GNC}}}$ and $\textsc{B}(x,y) = \int_0^1 t^{x-1}(1-t)^{y-1} dt$ corresponds to the Beta function (first order Euler function)~\cite{gradshteyn.07.integrals}.
\end{theorem}
\begin{IEEEproof}
Please refer to Appendix~\ref{ap:ps-csi-gnc}.
\end{IEEEproof}

From Theorem~\ref{th:ps-csi-gnc}, one can obtain the diversity order of the GNC scheme subjected to secrecy constraints as follows.
\begin{corollary} \label{corol:diversity-gnc}
The diversity order of the $M$-source GNC scheme with parameters $(k_1,k_2)$ is not reduced due to secrecy constraints, that is, it remains equal to $M+k_2$.
\end{corollary}
\begin{IEEEproof}
Please refer to Appendix~\ref{ap:diversity-gnc}.
\end{IEEEproof}
%
%
%

\subsection{Sources without CSI}

When CSI is not available at the transmitter side, the exact SOP of the two-source GNC is obtained according to~\eqref{eq:ps-nocsi2} after the substitution of $\mathcal{O}_{X}(\mathcal{R}, \bar{\gamma}_D)$ and $\mathcal{O}_{X}(\mathcal{R}_E, \bar{\gamma}_E)$ by the exact outage probability from~\eqref{eq:outage_gnc_exact}. Similarly to the case with partial CSI, in order to make the analysis more tractable, we favor E here by considering that the outage probability experienced by E is obtained under the assumption of perfect intersource channels, that is, $\mathcal{O}_{X}(\mathcal{R}_E, \bar{\gamma}_E)$ is dominated by the term  $\mathcal{O}_{1}(\mathcal{R}_E, \bar{\gamma}_E) $ in~\eqref{eq:outage_gnc_exact}. Thus, the SOP of a two-source network operating under the GNC scheme is approximated by
\begin{equation} \label{eq:ps-nocsi-gnc}
\begin{split}
\mathcal{S}_{\text{GNC}}^{\text{no-csi}} & \approx \mathcal{O}_{\text{GNC}}(\mathcal{R}, \bar{\gamma}_D) + \big[ 1-\mathcal{O}_{1}(\mathcal{R}_E, \bar{\gamma}_E)\big] \\
& \qquad - \mathcal{O}_{\text{GNC}}(\mathcal{R}, \bar{\gamma}_D)\big[ 1-\mathcal{O}_{1}(\mathcal{R}_E, \bar{\gamma}_E)\big],\\
\end{split}
\end{equation}
where $\mathcal{O}_{\text{GNC}}(\mathcal{R}, \bar{\gamma}_D)$ and $\mathcal{O}_{1}(\mathcal{R}_E, \bar{\gamma}_E)$ are obtained respectively from~\eqref{eq:outage_gnc_exact} and~\eqref{eq:outage_gnc_exact_p1}. Regarding its asymptotic behavior with the increase of $\bar{\gamma}_D$, the SOP from~\eqref{eq:ps-nocsi-gnc} presents a different behavior than the SOP of the GNC scheme with partial CSI from~\eqref{eq:ps-csi-gnc}, as presented in the following Theorem.
\begin{theorem} When the SNR of the legitimate nodes increases without limit, the SOP of the GNC scheme presents an outage floor which is given by
\begin{equation} \label{eq:ps-nocsi-gnc-se-asymptotic}
\begin{split}
\vec{\mathcal{S}}_{\text{GNC}}^{\text{no-csi}} &= \lim_{\bar{\gamma}_D \rightarrow \infty} \mathcal{S}_{\text{GNC}}^{\text{no-csi}} \\
& \approx 1-\mathcal{O}_{1}(\mathcal{R}_E, \bar{\gamma}_E).\\
\end{split}
\end{equation}
\end{theorem}
\begin{IEEEproof}
We can see from~\eqref{eq:outage_dt_rayleigh} that $\mathcal{O}(\mathcal{R}, \bar{\gamma}_D) \rightarrow 0$ when $\bar{\gamma}_D \rightarrow \infty$. By replacing  $\mathcal{O}(\mathcal{R}, \bar{\gamma}_D) = 0$ in~\eqref{eq:ps-nocsi2} and then in~\eqref{eq:ps-nocsi-gnc}, one can see that the SOP is limited by the secrecy outage event $(\mathcal{I}_{E}\geq \mathcal{R}_E)$, whose probability of occurrence for the $M$-source GNC scheme obtained from the complement of~\eqref{eq:outage_gnc_exact_free} 
leads to~\eqref{eq:ps-nocsi-gnc-se-asymptotic}, concluding the proof.
\end{IEEEproof}

In general, for the case of a $M$-source network, we can approximate the SOP from~\eqref{eq:ps-nocsi-gnc} by resorting to the high-SNR approximations of $\mathcal{O}_{\text{GNC}}(\mathcal{R}, \bar{\gamma}_D)$ and $\mathcal{O}_{1}(\mathcal{R}_E, \bar{\gamma}_E)$ given respectively in~\eqref{eq:outage_gnc} and~\eqref{eq:outage_gnc_free}, which leads to
\begin{equation} \label{eq:ps-nocsi-gnc-app}
\begin{split}
\tilde{\mathcal{S}}_{\text{GNC}}^{\text{no-csi}} & \approx 1\! -\! \mu_{\text{GNC}_1}\!\left[\!1\!-\!\exp\left(\!-\frac{2^{\mathcal{R}_E/R_{\text{GNC}}}-1}{\bar{\gamma}_E}\right) \! \right]^{Mk_2+1}  \\
& \times\!\left[\!1\!-\!\mu_{\text{GNC}_2}\!\left[1\!-\!\exp\left(\!-\frac{2^{\mathcal{R}/R_{\text{GNC}}}-1}{\bar{\gamma}_D} \right)\!\right]^{M+k_2} \right].
\end{split}
\end{equation}

However, as will be presented in the next section, even though the approximation from~\eqref{eq:ps-nocsi-gnc-app} is useful in obtaining an approximation to the slope of the SOP, it does not  accurately represent the floor in the SOP according to~\eqref{eq:ps-nocsi-gnc-se-asymptotic}. Thus, in what follows we present a more accurate approximation for the $M$-source GNC scheme than~\eqref{eq:ps-nocsi-gnc-app}.
\begin{corollary} \label{corol:ps-nocsi-app}
The SOP of the $M$-source GNC scheme can be approximated as
\begin{equation} \label{eq:ps-nocsi-gnc-app2}
\mathcal{S}_{\text{GNC}}^{\text{no-csi}} \approx \max\left\{\vec{\mathcal{S}}_{\text{GNC}}^{\text{no-csi}}, \tilde{\mathcal{S}}_{\text{GNC}}^{\text{no-csi}} \right\},
\end{equation}
 where $\vec{\mathcal{S}}_{\text{GNC}}^{\text{no-csi}}$ and $\tilde{\mathcal{S}}_{\text{GNC}}^{\text{no-csi}} $ are obtained from~\eqref{eq:ps-nocsi-gnc-se-asymptotic} and~\eqref{eq:ps-nocsi-gnc-app}, respectively.
\end{corollary}
\begin{IEEEproof}
Since the approximation~\eqref{eq:ps-nocsi-gnc-app} does not reflect the actual behavior of the floor in the SOP, one can limit the SOP to the aforementioned floor by taking the maximum between the approximation in~\eqref{eq:ps-nocsi-gnc-app} and the asymptotic value from~\eqref{eq:ps-nocsi-gnc-se-asymptotic}.
\end{IEEEproof}


\section{Numerical Results}
\label{sec:numerical_results}

\begin{figure} [!t]
\begin{center}
\includegraphics[width=0.5\textwidth]{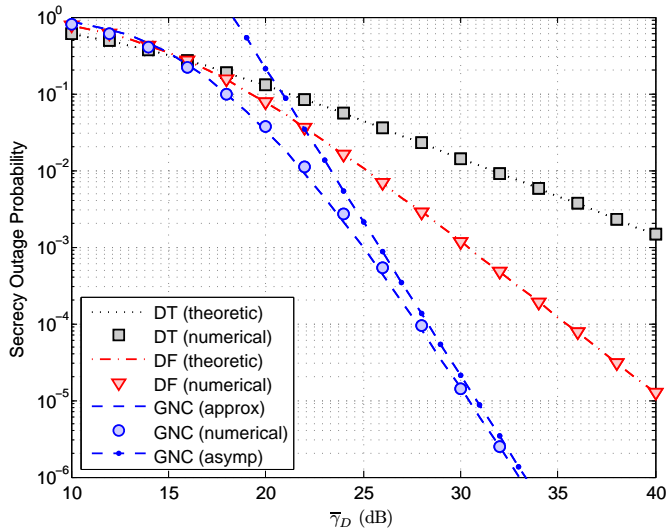}
\end{center}
\caption{Secrecy outage probability versus $\bar{\gamma}_{D}$ for the DT, DF and GNC schemes (with $M=k_1=k_2=2$), considering $\mathcal{R}_s=0.5$ bpcu and Eve's average SNR $\bar{\gamma}_{E} = 10$ dB. Partial CSI scenario.}
\label{fig:ps_csi_snr_rs05_snre10}
\end{figure}

\begin{figure}
\begin{center}
\includegraphics[width=0.5\textwidth]{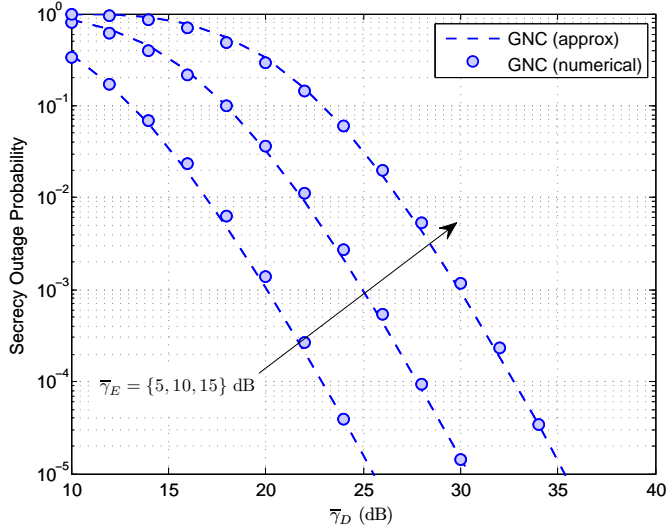}
\end{center}
\caption{Secrecy outage probability versus $\bar{\gamma}_{D}$ for GNC scheme (with $M=k_1=k_2=2$), considering $\mathcal{R}_s=0.5$ bpcu and Eve's average SNR $\bar{\gamma}_{E} = \{5,10,15\}$~dB. Partial CSI scenario.}
\label{fig:ps_csi_snr_rs05_snre51015}
\end{figure}

\begin{figure}
\begin{center}
\includegraphics[width=0.5\textwidth]{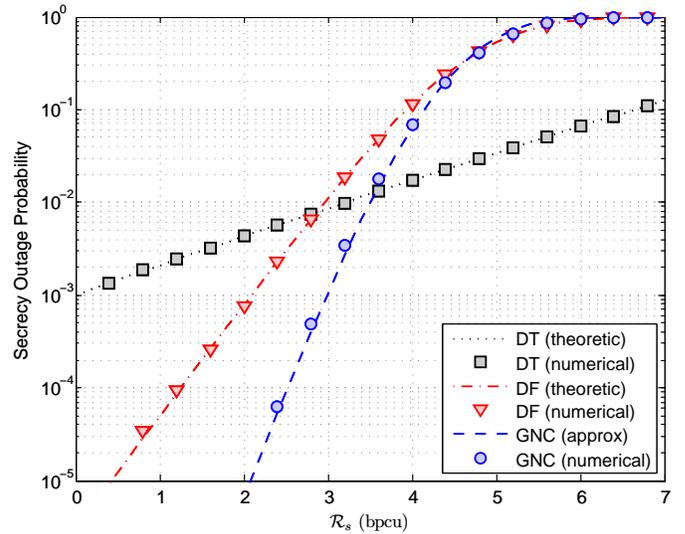}
\end{center}
\caption{Secrecy outage probability versus $\mathcal{R}_s$ for DT, DF and GNC (with $M=k_1=k_2=2$) schemes, considering $\bar{\gamma}_{D} = 40$~dB and $\bar{\gamma}_{E} = 10$~dB. Partial CSI scenario.}
\label{fig:ps_csi_rs_snr40_snre10}
\end{figure}

\begin{figure}
\begin{center}
\includegraphics[width=0.5\textwidth]{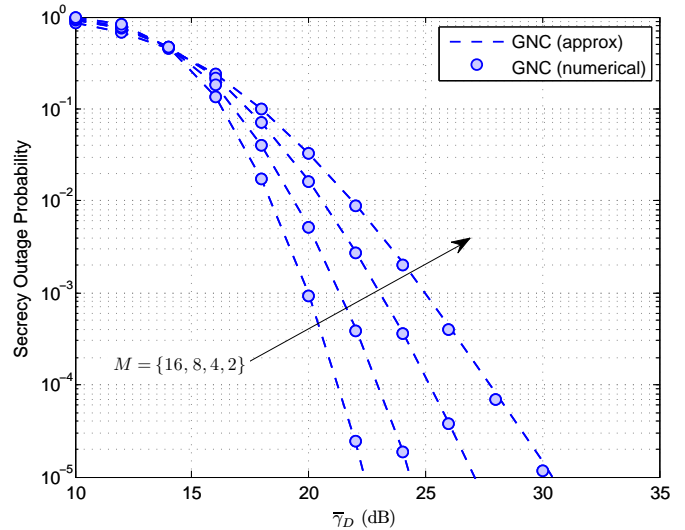}
\end{center}
\caption{Secrecy outage probability versus $\bar{\gamma}_{D}$ for GNC scheme (with $k_1=k_2=2$), considering $\mathcal{R}_s=0.5$ bpcu, $\bar{\gamma}_{E} = 10$~dB and $M = \{2,4,8,16\}$. Partial CSI scenario.}
\label{fig:ps_csi_snr_rs05_snre10_M24816}
\end{figure}

\begin{figure} [!t]
\begin{center}
\includegraphics[width=0.5\textwidth]{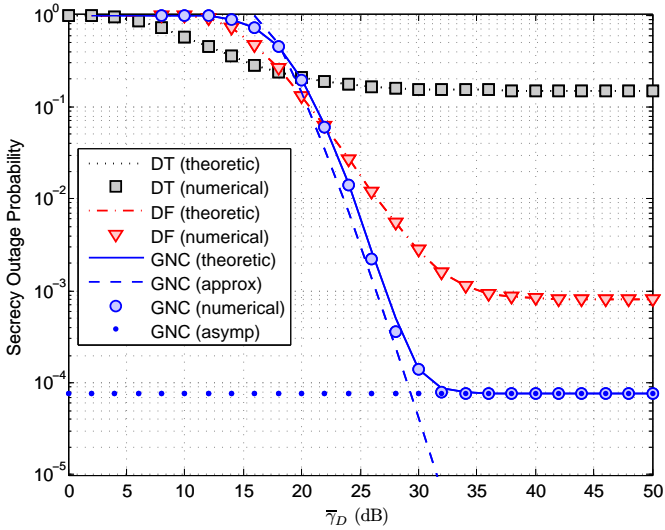}
\end{center}
\caption{Secrecy outage probability versus $\bar{\gamma}_{D}$ for the DT, DF and GNC (with $M=k_1=k_2=2$) schemes without CSI at the transmitters, considering $\mathcal{R}=3$ bpcu, $\mathcal{R}_E=2$ bpcu and Eve's average SNR $\bar{\gamma}_{E} = 2$~dB. No CSI scenario.}
\label{fig:ps_nocsi_snr_r3_re2_snre0}
\end{figure}

\begin{figure} [!t]
\begin{center}
\includegraphics[width=0.5\textwidth]{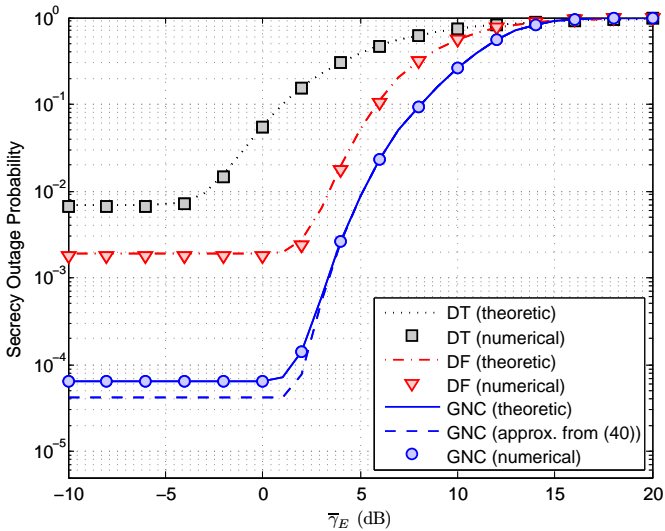}
\end{center}
\caption{Secrecy outage probability versus $\bar{\gamma}_{E}$ for the DT, DF and GNC (with $M=k_1=k_2=2$) schemes without CSI at the transmitters, considering $\mathcal{R}=3$ bpcu, $\mathcal{R}_E=2$ bpcu and $\bar{\gamma}_{D}~=~30$~dB. No CSI scenario.}
\label{fig:ps_nocsi_snre_r3_re2_snr30}
\end{figure}

\begin{figure} [!t]
\begin{center}
\includegraphics[width=0.5\textwidth]{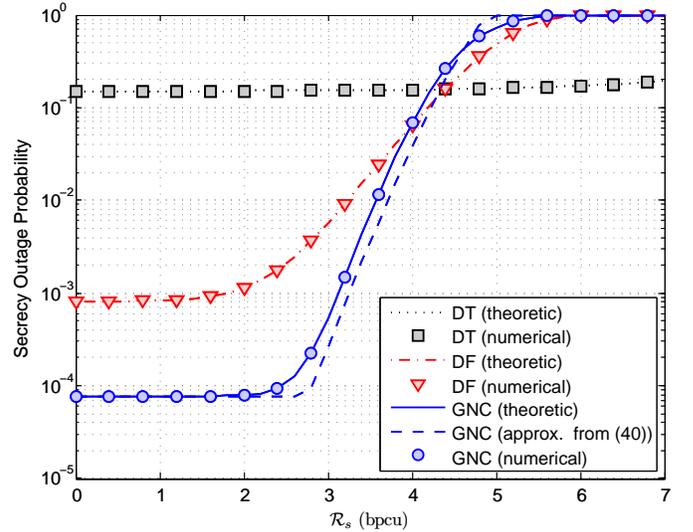}
\end{center}
\caption{Secrecy outage probability versus $\mathcal{R}_s$ for the DT, DF and GNC (with $M=k_1=k_2=2$) schemes, considering $\mathcal{R}_E=2$ bpcu, $\bar{\gamma}_{E}~=~2$~dB and $\bar{\gamma}_{D}~=~40$~dB. No CSI scenario.}
\label{fig:ps_nocsi_rs_re2_snr40_snre2}
\end{figure}

\begin{figure} [!t]
\begin{center}
\includegraphics[width=0.5\textwidth]{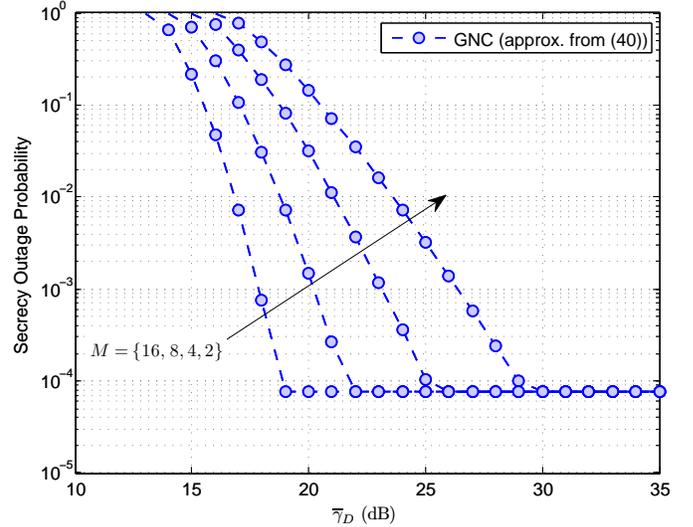}
\end{center}
\caption{Secrecy outage probability versus $\bar{\gamma}_{D}$ for the GNC scheme (with $k_1=k_2=2$), considering $\mathcal{R}=3$ bpcu, $\mathcal{R}_E=2$ bpcu, $\bar{\gamma}_{E}~=~2$~dB and $M~=~\{2,4,8,16\}$. No CSI scenario.}
\label{fig:ps_nocsi_snr_r3_re2_snre2_M24816}
\end{figure}

In this section, we present some numerical results in order to validate the results obtained analytically. For the GNC scheme, the instantaneous SNRs $\gamma_{D}$ and $\gamma_{E}$ were obtained according to the inverse transform sampling method~\cite{devroye.86}. In what follows, Figs.~\ref{fig:ps_csi_snr_rs05_snre10}-\ref{fig:ps_csi_snr_rs05_snre10_M24816} refer to the case with partial CSI at the sources, while the scenario without any CSI at the sources is evaluated in Figs.~\ref{fig:ps_nocsi_snr_r3_re2_snre0}-\ref{fig:ps_nocsi_snr_r3_re2_snre2_M24816}. Unless stated otherwise, we assume the parameters $M\!=\!k_1\!=\!k_2\!=\!2$ for the GNC scheme, in order to perform a fair comparison to the DF scheme, such that the number of users and the code rate are the same.

Fig.~\ref{fig:ps_csi_snr_rs05_snre10} presents the SOP versus $\bar{\gamma}_{D}$ for the DT, DF and GNC schemes, considering that $\mathcal{R}_s~=~0.5$ bits per channel use (bpcu) and that $\bar{\gamma}_{E}~=~10$~dB, as well as assuming CSI of the legitimate channels at the sources. For the GNC scheme in Fig.~\ref{fig:ps_csi_snr_rs05_snre10}, ``numerical'' refers to the numerical result obtained from the exact SOP via Monte Carlo method, while ``approx'' refers to the analytical high-SNR approximation from~\eqref{eq:ps-csi-gnc}. We can see that the GNC scheme presents the highest diversity order (slope of the curve) among all the three schemes, matching the value $M\!+\!k_2=4$ (represented by the curve ``asymp'') according to Corollary~\ref{corol:diversity-gnc}, which means that the GNC scheme outperforms the other schemes when a low SOP is required. It can also be seen that the numerical results match the analytical ones with good precision.

The influence of $\bar{\gamma}_{E}$ in the SOP performance of the GNC scheme with partial CSI is evaluated in Fig.~\ref{fig:ps_csi_snr_rs05_snre51015}, considering that $\mathcal{R}_s~=~0.5$~bpcu and $\bar{\gamma}_{E}~=~ \{5,10,15\}$~dB. We can see that when $\bar{\gamma}_{E}$ increases, the SOP performance is degraded in terms of coding gain (the curve is moved to the right). However, the diversity order remains unchanged.

Fig.~\ref{fig:ps_csi_rs_snr40_snre10} evaluates the SOP of the DT, DF and GNC schemes with partial CSI as a function of the secrecy rate $\mathcal{R}_s$. One can see that the cooperative schemes outperform the DT up to a certain threshold of $\mathcal{R}_s$, from which the DT presents the lowest SOP. However, for the low-SOP region, the GNC scheme is the one that presents the best performance.

 The SOP versus $\bar{\gamma}_{D}$ for the GNC scheme with different numbers of sources $M$ is presented in Fig.~\ref{fig:ps_csi_snr_rs05_snre10_M24816}. We can see that the performance of the scheme, as its diversity order, increases as the number of users increases. However, the relative gains diminish with the increase of $M$.

Fig.~\ref{fig:ps_nocsi_snr_r3_re2_snre0} considers the case without CSI at the sources, and shows the SOP versus $\bar{\gamma}_{D}$ for the situation where $\mathcal{R}=3$~bpcu, $\mathcal{R}_E=2$~bpcu and $\bar{\gamma}_{E} = 2$~dB. In Fig.~\ref{fig:ps_nocsi_snr_r3_re2_snre0}, ``theoretic'' refers to the SOP obtained from~\eqref{eq:ps-nocsi-gnc}, ``approx'' refers to the approximation in~\eqref{eq:ps-nocsi-gnc-app} and ``asymp'' to the asymptotic result according to~\eqref{eq:ps-nocsi-gnc-se-asymptotic}. One can see that all the considered schemes present an error floor which is limited by the secrecy outage event, however, the floor of the GNC scheme is the lowest. It is also worth noting that the exact SOP is well approximated by the maximum between the approximated and the asymptotic results from~\eqref{eq:ps-nocsi-gnc-app2}, Corollary~\ref{corol:ps-nocsi-app}.

The influence of $\bar{\gamma}_{E}$ in the SOP performance in the case without CSI is evaluated in Fig.~\ref{fig:ps_nocsi_snre_r3_re2_snr30}, for the DT, DF and GNC schemes with $\mathcal{R}~=~3$~bpcu, $\mathcal{R}_E~=~2$~bpcu and $\bar{\gamma}_{D}~=~30$~dB. One can see that the GNC scheme presents the lowest SOP for the whole considered range. It can also be seen that the approximation from~\eqref{eq:ps-nocsi-gnc-app2} is useful in obtaining an approximation for the SOP of the GNC scheme.

Fig.~\ref{fig:ps_nocsi_rs_re2_snr40_snre2} evaluates the influence of the secrecy rate $\mathcal{R}_s$ on the performance of the DT, DF and GNC schemes, considering that $\mathcal{R}_E=2$ bpcu, $\bar{\gamma}_{E}~=~2$~dB and $\bar{\gamma}_{D}~=~40$~dB. We can see that the SOP of all schemes increase when $\mathcal{R}_s$ increases. However, the GNC is the scheme that presents the lowest SOP in the region of low SOP.


Fig.~\ref{fig:ps_nocsi_snr_r3_re2_snre2_M24816} presents the SOP versus $\bar{\gamma}_{D}$ for the GNC scheme with different numbers of sources $M$. We can see that the performance of the scheme is improved as the number of users increases. However, besides the relative gains diminish with the increase of $M$, one can see that the floor in the SOP is approximately the same, independently of $M$. This is due to the fact that the floor corresponds to the complement of the outage probability so that its variation is only noticeable at the high SOP range.

\section{Final Comments}
\label{sec:final_comments}

We evaluated the secrecy outage probability of a multi-source cooperative network where the destination node is wiretapped by a malicious and passive eavesdropper. We proposed the application of the network coding technique as an alternative to increase the secrecy at the destination node. Different scenarios with partial and completely without channel state information (CSI) at the sources were evaluated and we showed through theoretic and numerical analyses that the secrecy can be significantly increased through the use of network coding when compared to the direct transmission and traditional cooperative techniques.

\appendices

\section{Proof of Theorem \ref{th:ps-csi-gnc}} \label{ap:ps-csi-gnc}

\begin{IEEEproof}
 After replacing the CDF of $\gamma_{D}$ obtained from~\eqref{eq:cdf_d} and the pdf of $\gamma_{E}$ obtained from~\eqref{eq:pdf_e} in~\eqref{eq:cs_rs_nc_def}, we have that:
 \begin{equation} \label{ps-csi-gnc-proof-1}
\begin{split}
\mathcal{S}_{\text{\emph{GNC}}}^{\text{csi}}   &= \int_0^{\infty} F_{\gamma_{D}}(\gamma_U)p_{\gamma_{E}}(\gamma_{E}) d\gamma_{E}\\
& \approx \frac{Mk_2\!+\!1 }{\bar{\gamma}_{E}} \int_0^{\infty} \overbrace{\left[1-\exp\left(-\frac{\gamma_{U}}{\bar{\gamma}_{D}}\right)\right]^{M+k_2}}^{\textbf{A}} \\ &  \quad \times\exp\left(-\frac{\gamma_{E}}{\bar{\gamma}_{E}}\right)\left[1-\exp\left(-\frac{\gamma_{E}}{\bar{\gamma}_{E}}\right)\right]^{Mk_2} d\gamma_E.
\end{split}
\end{equation}
 We then expand term $\textbf{A}$ in~\eqref{ps-csi-gnc-proof-1} using the binomial expansion $\displaystyle [1-x]^n = \sum_{i=0}^n {n \choose i} [-1]^i x^i$, leading to:
  \begin{equation} \label{ps-csi-gnc-proof-2}
\begin{split}
\mathcal{S}_{\text{\emph{GNC}}}^{\text{csi}}   &\approx \frac{Mk_2\!+\!1  }{\bar{\gamma}_{E}} \int_0^{\infty} \sum_{i=0}^{M\!+\!k_2} {M\!+\!k_2 \choose i} [-1]^i \exp\left(-\frac{\gamma_{U}}{\bar{\gamma}_{D}}i\right) \\ &  \quad \times\exp\left(-\frac{\gamma_{E}}{\bar{\gamma}_{E}}\right)\left[1-\exp\left(-\frac{\gamma_{E}}{\bar{\gamma}_{E}}\right)\right]^{Mk_2}d\gamma_E.
\end{split}
\end{equation}
 According to~\cite[eq. (3.312.1)]{gradshteyn.07.integrals}, one has that:
  \begin{equation} \label{ps-csi-gnc-proof-3}
\int_{0}^{\infty} \!\left[1\!- \!\exp\left(\!-\frac{x}{\beta}\right)\right]^{\nu-1}\!\! \exp\left(-\alpha x\right) dx = \beta \textsc{B}\left(\beta \alpha, \nu\right).
 \end{equation}
 Thus, by substituting $\gamma_U=\xi(1+\gamma_{E})-1$ in~\eqref{ps-csi-gnc-proof-2} and substituting in~\eqref{ps-csi-gnc-proof-3} $\beta = \bar{\gamma}_E$, $\nu=Mk_2\!+\!1$, and $\alpha = (\xi\bar{\gamma}_E i + \bar{\gamma}_D)/(\bar{\gamma}_D\bar{\gamma}_E )$, we can obtain~\eqref{eq:ps-csi-gnc}, concluding the proof.

\end{IEEEproof}

\section{Proof of Corollary~\ref{corol:diversity-gnc}} \label{ap:diversity-gnc}

\begin{IEEEproof}
According to the definition of diversity order presented in~\eqref{eq:diversity_oder}, one must evaluate the asymptotically behavior of the outage probability with respect to $\bar{\gamma}_D$. For large values of $\bar{\gamma}_D$, the Beta function from~\eqref{eq:ps-csi-gnc} can be approximated as~\cite{gradshteyn.07.integrals}:
\begin{equation} \label{eq:beta-app}
\begin{split}
\textsc{B}\left(\frac{\xi\,\bar{\gamma}_{E}i + \bar{\gamma}_{D} }{\bar{\gamma}_{D}},Mk_2\!+\!1\right) & \approx \textsc{B}\left(1,Mk_2\!+\!1\right) \\
& = \frac{\Gamma(1)\,\Gamma(Mk_2\!+\!1)}{\Gamma(Mk_2\!+\!2)}\\
& = \frac{1}{Mk_2\!+\!1},
\end{split}
\end{equation}
where $\Gamma(\cdot)$ corresponds to the complete Gamma function~\cite{gradshteyn.07.integrals}. Thus, by replacing~\eqref{eq:beta-app} in~\eqref{eq:ps-csi-gnc}, one have that:
\begin{equation} \label{eq:ps-csi-gnc2}
\begin{split}
\mathcal{S}_{\text{\emph{GNC}}}^{\text{csi}}   &\approx \big[Mk_2\!+\!1\big]\!\!\sum_{i=0}^{M\!+\!k_2}\! {M\!+\!k_2 \!\choose\! i} [-1]^i \exp\!\left(\!-\frac{\xi-1}{\bar{\gamma}_{D}}i\right)  \\
& \quad \times \quad \left[\frac{1}{Mk_2\!+\!1}\right]\\
& = \sum_{i=0}^{M\!+\!k_2}\! {M\!+\!k_2 \!\choose\! i} [-1]^i \exp\!\left(\!-\frac{\xi-1}{\bar{\gamma}_{D}}i\right).
\end{split}
\end{equation}

After resorting to the fact that $\displaystyle \sum_{i=0}^n {n \choose i} [-1]^i x^i = \displaystyle [1-x]^n$, the SOP from~\eqref{eq:ps-csi-gnc2} can be rewritten as
\begin{equation} \label{eq:ps-csi-gnc3}
\mathcal{S}_{\text{\emph{GNC}}}^{\text{csi}}   \approx \left[1 - \exp\!\left(-\frac{\xi-1}{\bar{\gamma}_{D}}\right)\right]^{M\!+\!k_2}.
\end{equation}

Finally, the diversity order $M\!+\!k_2$ from Corollary~\ref{corol:diversity-gnc} is obtained by applying~\eqref{eq:ps-csi-gnc3} in~\eqref{eq:diversity_oder}.

\end{IEEEproof}

\bibliographystyle{ieeetran}
\bibliography{IEEEabrv,biblio}

\begin{thebibliography}{10}
\providecommand{\url}[1]{#1}
\csname url@samestyle\endcsname
\providecommand{\newblock}{\relax}
\providecommand{\bibinfo}[2]{#2}
\providecommand{\BIBentrySTDinterwordspacing}{\spaceskip=0pt\relax}
\providecommand{\BIBentryALTinterwordstretchfactor}{4}
\providecommand{\BIBentryALTinterwordspacing}{\spaceskip=\fontdimen2\font plus
\BIBentryALTinterwordstretchfactor\fontdimen3\font minus
  \fontdimen4\font\relax}
\providecommand{\BIBforeignlanguage}[2]{{%
\expandafter\ifx\csname l@#1\endcsname\relax
\typeout{** WARNING: IEEEtran.bst: No hyphenation pattern has been}%
\typeout{** loaded for the language `#1'. Using the pattern for}%
\typeout{** the default language instead.}%
\else
\language=\csname l@#1\endcsname
\fi
#2}}
\providecommand{\BIBdecl}{\relax}
\BIBdecl

\bibitem{shannon.49.secrecy}
C.~E. Shannon, ``Communication theory of secrecy systems,'' \emph{{Bell Syst.
  Tech. J.}}, vol.~28, pp. 656--715, 1949.

\bibitem{wyner.1975}
A.~D. Wyner, ``The wire-tap channel,'' \emph{{Bell Syst. Tech. J.}}, vol.~54,
  no.~8, pp. 1355--1387, 1975.

\bibitem{gopala.08}
P.~K. Gopala, L.~Lai, and H.~El-Gamal, ``On the secrecy capacity of fading
  channels,'' \emph{{IEEE} Trans. Inf. Theory}, vol.~54, no.~10, pp.
  4687--4698, Oct 2008.

\bibitem{barros.06.secrecy}
J.~Barros and M.~R.~D. Rodrigues, ``Secrecy capacity of wireless channels,'' in
  \emph{Proc. of the IEEE Int. Symp. on Inform. Theory (ISIT'06)}, 2006.

\bibitem{bloch.11}
M.~Bloch and J.~Barros, \emph{Physical-Layer Security: From Information Theory
  to Security Engineering}, C.~U. Press, Ed.\hskip 1em plus 0.5em minus
  0.4em\relax Cambridge University Press, 2011.

\bibitem{tang.07}
X.~Tang, R.~Liu, and P.~Spasojevic, ``On the achievable secrecy throughput of
  block fading channels with no channel state information at transmitter,'' in
  \emph{41st Annual Conference on Information Sciences and Systems, 2007. (CISS
  '07).}, March 2007, pp. 917--922.

\bibitem{tang.09}
X.~Tang, R.~Liu, P.~Spasojevic, and H.~Poor, ``On the throughput of secure
  hybrid-arq protocols for gaussian block-fading channels,'' \emph{{IEEE}
  Trans. Inf. Theory}, vol.~55, no.~4, pp. 1575--1591, April 2009.

\bibitem{alves.12.secrecy.tas}
H.~Alves, R.~D. Souza, M.~Debbah, and M.~Bennis, ``Performance of transmit
  antenna selection physical layer security schemes,'' \emph{IEEE Signal
  Process. Lett.}, vol.~19, no.~6, pp. 372--375, June 2012.

\bibitem{pheelep.13}
N.~Yang, P.~L. Yeoh, M.~Elkashlan, R.~Schober, and I.~B. Collings, ``Transmit
  antenna selection for security enhancement in {MIMO} wiretap channels,''
  \emph{{IEEE} Trans. Commun.}, vol.~61, no.~1, pp. 144--154, January 2013.

\bibitem{gabry.12}
F.~Gabry, ``Cooperation for secrecy in wireless networks,'' Ph.D. dissertation,
  KTH, School of Electrical Engineering, Communication Theory Laboratory,
  September 2012.

\bibitem{lai.08}
L.~Lai and H.~E. Gamal, ``The relay-eavesdropper channel: Cooperation for
  secrecy,'' \emph{{IEEE} Trans. Inf. Theory}, vol.~54, no.~9, pp. 4005--4019,
  September 2008.

\bibitem{kaido.14}
R.~Kaido, O.~K. Rayel, J.~L. Rebelatto, and R.~D. Souza, ``Network coded
  cooperation for a two-user wiretap channel,'' in \emph{Proc. of the IEEE Int.
  Conf. on Acoustics, Speech and Signal Processing (ICASSP'14)}, Florence,
  Italy, May 2014.

\bibitem{laneman.04}
J.~N. Laneman, D.~N.~C. Tse, and G.~W. Wornell, ``Cooperative diversity in
  wireless networks: Efficient protocols and outage bahavior,'' \emph{{IEEE}
  Trans. Inf. Theory}, vol.~50, no.~12, pp. 3062--3080, December 2004.

\bibitem{sendonaris.03}
A.~Sendonaris, E.~Erkip, and B.~Aazhang, ``User cooperation diversity: {P}art
  {I} and {P}art {II},'' \emph{{IEEE} Trans. Commun.}, vol.~51, no.~11, pp.
  1927--1948, November 2003.

\bibitem{meulen.71}
E.~C. van~der Meulen, ``Three-terminal communication channels,''
  \emph{{Advanced Applied Probability}}, vol.~3, pp. 120--154, 1971.

\bibitem{ahlswede.00}
R.~Ahlswede, N.~Cai, S.-Y. Li, and R.~Yeung, ``Network information flow,''
  \emph{{IEEE} Trans. Inf. Theory}, vol.~46, no.~4, pp. 1204 -- 1216, 2000.

\bibitem{xiao.10}
M.~Xiao and M.~Skoglund, ``Multiple-user cooperative communications based on
  linear network coding,'' \emph{{IEEE} Trans. Commun.}, vol.~58, no.~12, pp.
  3345--3351, December 2010.

\bibitem{rebelatto.10.TIT}
J.~L. Rebelatto, B.~F. Uchôa-Filho, Y.~Li, and B.~Vucetic, ``Multi-user
  cooperative diversity through network coding based on classical coding
  theory,'' \emph{{IEEE} Trans. Signal Process.}, vol.~60, no.~2, pp. 916--926,
  February 2012.

\bibitem{goldsmith.05}
A.~Goldsmith, \emph{Wireless Communications}.\hskip 1em plus 0.5em minus
  0.4em\relax Cambridge University Press, 2005.

\bibitem{gradshteyn.07.integrals}
I.~Gradshteyn and I.~M. Ryzhik, \emph{Table of Integrals, Series, and
  Products}, 7th~ed., A.~Jeffrey and D.~Zwillinger, Eds.\hskip 1em plus 0.5em
  minus 0.4em\relax Academic Press - Elsevier, 2007.

\bibitem{devroye.86}
L.~Devroye, \emph{Non-Uniform Random Variate Generation}, Springer-Verlag,
  Ed.\hskip 1em plus 0.5em minus 0.4em\relax New York: Springer-Verlag, 1986.

\end{thebibliography}

\end{document}